\documentclass[12pt]{article}
\usepackage{epsfig}
\usepackage{amssymb}
\usepackage{pslatex}

\makeatletter

\@addtoreset{equation}{section}
\makeatother
\newcommand{\bea}{\begin{eqnarray}}
\newcommand{\eea}{\end{eqnarray}}
\newcommand{\be}{\begin{eqnarray}}
\newcommand{\ee}{\end{eqnarray}}

\def\fR{{\mathfrak R}}
\def\fL{{\mathfrak L}}
\def\fQ{{\mathfrak Q}}
\def\fS{{\mathfrak S}}

\def\rmd{{\rm d}}

\begin{document}

\begin{titlepage}
\vskip-1cm
\begin{flushright}
{\tt arXiV:0904.4677}\\
SNUST {\tt 090401}\\
UOSTP {\tt 09031}
\end{flushright}
\vskip0.5cm
\centerline{\Large \bf Generalized Dynamical Spin Chain and 4-Loop Integrability}
\vskip0.25cm
\centerline{\Large \bf in}
\vskip0.25cm
\centerline{\Large \bf ${\cal N}=6$ Superconformal Chern-Simons Theory }
\vskip1.25cm
\centerline{\large  Dongsu Bak $^{a}$, Hyunsoo Min $^{a}$, Soo-Jong Rey $^{b,c}$
}
\vspace{1cm}
\centerline{\sl a) Physics Department, University of Seoul, Seoul
130-743 {\rm KOREA}}
\vspace{0.25cm}
\centerline{\sl b) School of Physics and Astronomy, Seoul National University, Seoul 151-747 {\rm KOREA}}
\vspace{0.25cm}
\centerline{\sl c) Kavli Institute for Theoretical Physics, University of California, Santa Barbara
93106 {\rm USA}}
\vspace{0.50cm}
\centerline{\tt dsbak@uos.ac.kr,\,\,\, hsmin@dirac.uos.ac.kr, \,\,\, sjrey@snu.ac.kr }
\vspace{1cm}
\centerline{\bf abstract}
\vspace{0.75cm}
\noindent
We revisit unitary representation of centrally extended $\mathfrak{psu}(2 \vert 2)$ excitation superalgebra. We find most generally that `pseudo-momentum', not lattice momentum, diagonalizes spin chain Hamiltonian and leads to generalized dynamic spin chain. All known results point to lattice momentum diagonalization for ${\cal N}=4$ super Yang-Mills theory. Having different interacting structure, we ask if ${\cal N}=6$ superconformal Chern-Simons theory provides an example of pseudo-momentum diagonalization. For SO(6) sector, we study maximal shuffling and next-to-maximal shuffling terms in the dilatation operator and compare them with results expected from $\mathfrak{psu}(2 \vert 2)$ superalgebbra and integrability. At two loops, we rederive maximal shuffling term (3-site) and find perfect agreement with known results. At four loops, we first find absence of next-to-maximal shuffling term (4-site), in agreement with prediction based on integrability. We next extract maximal shuffling term (5-site), the most relevant term for checking the possibility of pseudo-momentum diagonalization. Curiously, we find that result agrees with integraility prediction based on lattice momentum, as in ${\cal N}=4$ super Yang-Mills theory. Consistency of our results is fully ensured by checks of renormalizability up to six loops.

\end{titlepage}

\section{Introduction}
The holographic duality between Type IIA string theory on AdS$_4 \times \mathbb{CP}^3$ and (2+1)-dimensional ${\cal N}=6$ superconformal Chern-Simons theory, discovered by Aharony, Bergman, Jafferis and Maldacena (ABJM) \cite{Aharony:2008ug}, opened a new avenue for exploring the AdS/CFT correspondence \cite{Maldacena:1997re}. The (2+1)-dimensional superconformal field theory admits Lagrangian formulation, hereafter referred as ABJM theory, so a systematic study between bulk fields and boundary operators is feasible much to the same extent as the holographic duality between Type IIB string theory on AdS$_5 \times \mathbb{S}^5$ and (3+1)-dimensional ${\cal N}=4$ super Yang-Mills (SYM) theory. The ABJM theory depends on two parameters, level $k$ of the Chern-Simons term and rank $N$ of the gauge group. The planar limit, $N \rightarrow \infty$, of the ABJM theory corresponds to non-interacting limit, $g_s \rightarrow 0$, of the Type IIA string theory. In this limit, the ABJM theory is organized in perturbation theory at weak `t Hooft coupling regime $\lambda \equiv (N / k) \ll 1$, while the Type IIA string theory is organized in perturbation theory at strong `t Hooft coupling regime $\sqrt{\lambda} \gg 1$.

A very interesting and important aspect of the AdS$_4$-ABJM holographic duality is the prospect of `integrability', the feature investigated thoroughly in the AdS$_5-$SYM holographic duality~ \cite{Minahan:2002ve}-\cite{Beisert:2006qh}.
At strong `t Hooft coupling regime, Type IIA string worldsheet dynamics was found integrable at leading order in $1/\sqrt{\lambda}$ \cite{classicalintegrability, Bak}. At weak `t Hooft coupling regime, SO(6) sector of the ABJM dilatation operator was found integrable at 2 loops \cite{Mina, Bak}. Extension to full OSp$(6|4)$ dilatation operator and Bethe ansatz equations were proposed to all orders and numerous consistency checks of the proposal were studied \cite{allloop} - \cite{gm4}. It was found that the excitation spectrum in the AdS$_4$-ABJM theory is organized by a direct sum of $\mathfrak{psu}(2 \vert 2)$ superalgebra. Interestingly, a direct product of the same superalgebra also featured in the excitation spectrum in the AdS$_5$-SYM theory \cite{Beisert:2005tm}, \cite{Beisert:2006qh}. These results all point toward exact `quantum integrability' of the ABJM theory, but a direct analytic proof would be highly desirable.

The purpose of this paper is to study the dilatation operator of the ABJM theory beyond leading order. Our motivations are primarily twofold. First, we wish to understand precise dynamic nature of the alternating spin chain in this theory. In the case of ${\cal N}=4$ super Yang-Mills theory, dilatation operator was described by a novel spin chain whose length changes dynamically \cite{Beisert:2003ys}. This novelty was in fact a simple consequence of $\mathfrak{psu}(2,2|4)$ superconformal algebra. For the $\mathfrak{osp}(6|4)$ superconformal algebra of the ABJM theory, we also expect emergence of dynamic spin chain. Magnon excitations in both theories are known to be described by product or sum of {\sl centrally-extended} $\mathfrak{psu}(2 \vert 2)$. On the other hand, field contents and their interactions as well as supersymmetry preserved by the spin chain ground-state differ for the two theories. Therefore, precise dynamic nature of the spin chain could also differ each other. Second, we would like to test quantum integrablity of the operator contents in this theory. While there were several indications from both weak and strong coupling regime, there was no rigorous proof yet. We address this problem by comparing
spin chain Hamiltonian expected from integrability and weak coupling expansion of it with the dilatation operator computed directly from the ABJM theory at lower but beyond leading orders in perturbation theory.

We begin in section 2 with centrally extended $\mathfrak{psu}(2|2)$ superalgebra, the superalgebra that features both the ABJM and the ${\cal N}=4$ SYM theories. Re-examining the previous study carefully, we find that off-shell magnon excitations, described by unitary representations of the superalgebra, is in general labeled by a `pseudo-momentum' which in general differs from the lattice momentum of the spin chain. Functional relation of the `pseudo-momentum' to the lattice momentum depends on the theory under consideration, on the amount of supersymmetry and order of perturbative interactions. In turn, the `pseudo-momentum' different from the lattice momentum lead to a generalized dynamical spin chain in that the exchange algebra of the magnon excitation is generalized order by order in perturbation theory.

In section 3, focusing on a magnon in SU(2)$\oplus$SU(2) sector, we compare predictions of quantum integrability to the spectrum of operator contents, equivalently, to general structure of the spin chain Hamiltonian at higher orders in perturbation theory.
We focus in the dilatation operator on so-called `maximal shuffling' terms - terms that exchange spins at furthest sites - and `next-to-maximal shuffling' terms - terms that exchange spins one less than furthest sites - at each order, since these terms are the cleanest to compute yet highly nontrivial in structure.
We explain diagrammatic origin of these terms up to four loops.
At two loops, we rederive maximal shuffling term (3-sites) and find perfect agreement with known results. At four loops, we first find absence of next-to-maximal shuffling term (4-sites), in agreement with integrability. We then extract maximal shuffling term (5-sites).

In section 4, we explain formalism for computing anomalous dimensions from two-point function of single-trace operators in perturbation theory and hence the quantum dilatation operator. We also explain general structure of the dilatation operators and how the maximal shuffling terms of length $2\ell+1$ and next-to-maximal shuffling terms of order $2 \ell$ arise diagrammatically at $2\ell$-th order in perturbation theory.
In sections 5 and 6, we compute quantum dilatation operator in dimensional regularization. In section 5, we first compute two-loop contribution to two-point function and reproduce the known result of the dilatation operator. We then compute four-loop contribution. Combining both contributions, we extract next-to-maximal
shuffling terms at four loops. If quantum integrability holds, result from section 3 indicates these terms should be absent. We indeed find that bosonic and fermionic loop contributions cancel each other -- a clear
indication that integrability holds. In section 6, we focus on maximal shuffling terms at four loops, where only bosonic diagrams contribute. Again, we find the result agrees with the prediction of integrability based on lattice momentum in section 3. Internal consistency of our results is fully ensured by checks of renormalizability up to six loops.

Along with ${\cal N}=4$ SYM theory, our result escalates a puzzle why the dilatation oprators are always diagonalized by lattice momentum eigenstates, not by pseudo-momentum eigenstates. In section 7, we discuss implication of this to other issues pertinent to the ABJM theory. We also discuss various implications of our results. In Appendix A, we present $\mathfrak{psu}(2 \vert 2)$ S-matrices in the pseudo-momentum basis. Appendix B is relegated to technical details of precision numerical evaluation of irreducible Feynman diagrams, whose results were used in sections 5 and 6.

\section{Off-Shell $\mathfrak{PSU}(2 \vert 2)$ and Generalized Dynamic Spin Chain}


In the planar limit, we consider a single trace operator of the ABJM scalar fields. In ABJM theory, the
scalar fields $Y^I, Y^\dagger_I$ $(I=1,2,3,4)$ transform as $({\bf N}, \overline{\bf N}; {\bf 4})$ and
$(\overline{\bf N}, {\bf N}; \overline{\bf 4})$ under the U$(N)\times\overline{\rm U}(N)$ gauge group and
the SU(4) R-symmetry group. Consider in the planar limit single-trace operators ${\cal O}$. By gauge invariance, they form an alternating spin chain, labeled by SU(4) quantum numbers:
\be
{\cal O}\,[I_1\, I_2 \, I_3 \, I_4 \, I_5 \,\cdots, I_{2L} ]={\rm Tr}( \, Y^{I_1}\,
Y^\dagger_{I_2}\,Y^{I_3}\, Y^\dagger_{I_4}\, Y^{I_5}\,\cdots\, Y^\dagger_{I_{2L}} )\,.
\ee
In the limit $L \rightarrow \infty$, choose the convention that $Y^I$ fields of ${\bf 4}$ are in the odd sites of the spin chain and $Y^\dagger_I$ of $\overline{\bf 4}$ in the even sites. By group theory, ${\bf 4}$ at odd sites and $\overline{\bf 4}$ at even sites interact only through $\delta^I_J$.

Rename $(Y^1, Y^2, Y^3, Y^4)$ as $(A_1, A_2, B^\dagger_2, B^\dagger_1)$ where $A_a, B_b$ ($a,b=1,2$) transform under the SU(2)$_A$ and the SU(2)$_B$ subgroups of SU(4) \cite{su2}. We focus on a restricted set of single-trace operators where $A_a$ and $\,\,B_b$ field are placed at odd/$\,\,$even sites, respectively:
\be
O\,[\,a_1 \,a_2\, a_3\, a_4\, a_5 \,\cdots ]={\rm Tr}(\, A_{a_1}\,
B_{a_2}\,A_{a_3}\, B_{a_4}\, A_{a_5}\,\cdots\,)\,.
\ee
A given operator of this type is mappable to a given spin chain state $|\, a_1 \, a_2 \, a_3 \cdots \rangle$. We shall take the ground state as
\bea
|O\rangle = \vert \, 1 \, 1 \, 1 \, 1  \cdots \rangle \qquad
\leftrightarrow \qquad
O\,[\, 1\, 1\, 1\, 1 \cdots ] =  {\rm Tr}(\, A_{1}\,
B_{1}\,A_{1}\, B_{1}\, A_{1}\, B_{1}\,\cdots\,)
\,.
\label{vacuum}
\eea
For this type of spin chains, there is no interaction between ${\bf 4}$ and $\overline{\bf 4}$ spins. Hence, the odd-site spin chain and the even-site spin chain behave independently. Then, there will be $A$-type magnons and $B$-type magnons propagating independently without interactions between the two types
\footnote{In principle, interaction of double-exchange $\mathbb{P}_{n+1,n+3}\, \mathbb{P}_{n+2, n+4}$ could arise, where $\mathbb{P}_{ij}$ denotes the permutation operator exchanging spins at the $i$-th and the $j$-th sites. In this case, there will be interactions between these two types of magnons.}.

The magnon excitations are organized by off-shell $\mathfrak{psu}(2 \vert 2) \oplus \mathfrak{psu}(2 \vert 2)$ superalgebra symmetries acting on $A$- and $B$-sites, respectively. The $({\bf 2} \vert {\bf 2})$ excitation multiplets are $(A_2, B_2^\dagger \vert \Psi_1, \Psi_2)$ and $(B_2, A_2^\dagger \vert \Psi_2^\dagger, \Psi_1^\dagger)$. Therefore, magnons of $A$- and $B$-types behave identical. From now on, we shall focus on $A$-type magnons and denote $(A_2, B_2^\dagger \vert \Psi_1, \Psi_2)$ as $\Phi = (\phi^1, \phi^2 \vert \psi^1, \psi^2)$ collectively. One expects a magnon excitation is an eigenstate of lattice momentum of the $A$-type sites in the alternating spin chain:
\bea
|p \rangle =\sum_{n=0}^{L-1}
e^{i n p}\,|\cdots \overbrace{\Phi}^{2n+1} \cdots\rangle \, . \label{latticemomentumeigenstate}
\eea
Here, the ellipses denote spin configurations of the ground state, where $A_1$ or $B_1$ in the remaining odd or even sites. Below, we shall suppress denoting these background spins explicit.

The off-shell $\mathfrak{psu}(2|2)$ superalgebra of the excitation symmetry is spanned by the two su(2) rotation generators ${\fR}^a\!_b$, ${\fL}^\alpha\!_\beta$, the supersymmetry generator ${\fQ}^\alpha_a$ and the superconformal generator ${\fS}^a_\alpha$. The off-shell configuration is characterized by $sl(2, \mathbb{R})$ central charges $\mathfrak{C}, \mathfrak{K}, \mathfrak{K}^*$~\cite{Beisert:2005tm}.
Their (anti)commutators are  given by \cite{Beisert:2005tm}
\bea
&& [{\fR}^a\!_b, \,\, \mathfrak{J}^c]= \delta^c_b \,
\mathfrak{J}^a-{1\over 2}
\delta^a_b \, \mathfrak{J}^c\,,\ \ \
[{\fL}^\alpha\!_\beta, \,\, \mathfrak{J}^\gamma]= \delta^\gamma_\beta  \,
\mathfrak{J}^\alpha-{1\over 2}
\delta^\alpha_\beta \, \mathfrak{J}^\gamma\,
\nonumber\\
&& \{ {\fQ}^\alpha_a, \,\, \fS^b_\beta\}= \delta^b_a   {\fL}^\alpha\!_\beta
+ \delta^\alpha_\beta \, {\fR}^b\!_a
+ \delta^b_a  \delta^\alpha_\beta \mathfrak{C}            \nonumber \\
&& \{ {\fQ}^\alpha_a, \,\, \fQ^\beta_b\}=\epsilon^{\alpha\beta}
\epsilon_{ab} \mathfrak{K}\,,\ \ \
\{ {\fS}^a_\alpha, \,\, \fS_\beta^b\}=\epsilon_{\alpha\beta}
\epsilon^{ab} \mathfrak{K}^* \,.
\eea
The central charges $\mathfrak{C}$ is related to the energy
by $E= \mathfrak{C}$, while $\mathfrak{K}, \mathfrak{K}^*$ introduced at off-shell are related to the momentum.
On a state of fundamental representation, the generators act as
\bea
&& \fR^a\!_b |\phi^c \rangle =
\delta^c_b  |\phi^a \rangle
-{1\over 2} \delta^a_b  |\phi^c\rangle
\,, \ \ \
\fL^\alpha\!_\beta |\phi^\gamma \rangle =
\delta^\gamma_\beta  |\phi^\alpha \rangle
-{1\over 2} \delta^\alpha_\beta  |\phi^\gamma\rangle
\eea
and as
\bea
&&\fQ^\alpha_a |\phi^b \rangle \, = \,
a \, \delta_a^b  |\psi^\alpha \rangle \nonumber \\
&& \fQ^\alpha_a |\psi^\beta \rangle =\,
b \, \epsilon^{\alpha\beta}\epsilon_{ab}  |\phi^b G(A^+) \rangle\nonumber\\
&&\fS^a_\alpha|\phi^b \rangle \, =\,
c \, \epsilon_{\alpha\beta}\epsilon^{ab} |\psi^\beta {G}({A^-})
\rangle \nonumber \\
&& \fS^a_\alpha |\psi^\beta \rangle =\,
d \, \delta^\beta_{\alpha}  |\phi^a \rangle\,.
\label{repre}
\eea
The function $G(A^+)$ is not arbitrary. Note first that $A^+$ inserts the ground-state spin $A_1$ into the odd site of the chain while $A^-$ removes one ground-state spin $A_1$.
Acting on an eigenstate (\ref{latticemomentumeigenstate}) of the lattice momentum $p$, we have
\bea
G(A^\pm) = A^\pm \qquad \mbox{where} \qquad |A^\pm \Phi \rangle = e^{\mp ip} | \Phi A^\pm \rangle \, .
\eea
This is the defining relation of the new marker function $G(A^\pm)$.

Closure of the superalgebra on fundamental representation leads to the `shortening condition'
\bea
a d-b c=1 \, . \label{shortening1}
\eea
One also finds the central charges of total momentum yield
\bea
&& \mathfrak{K}\,\,\, |\, \Phi \, \rangle  = a b\,
|\Phi \, G(A^+) \rangle
 \nonumber \\ 
&& \mathfrak{K}^* |\, \Phi \, \rangle  = c d\,
|\, \Phi \, {G}(A^-) \, \rangle \label{kkstar}
\eea
and the central charge of total dilatation energy yields
\bea
\mathfrak{C} \,\,\, |\, \Phi \, \rangle = {1 \over 2} (ad + bc) |\, \Phi \, \rangle. \label{dilatation}
\eea

Quantum mechanically, functional form of $G(A^\pm)$ in (\ref{repre}) are subject to radiative corrections. These corrections can be extracted straightforwardly, for example, from explicit derivation of quantum dilatation operator and utilization of the relations (\ref{kkstar}, \ref{dilatation}) or from operator product expansions between fermionic charges $\mathfrak{Q}, \mathfrak{S}$ and a pair of single-trace operators carrying bosonic and fermionic excitations, respectively.
After radiative corrections are taken into account, the full-fledged quantum marker function $G(A^\pm)$ ought to solve the `shortening condition' (\ref{shortening1}). It is an interesting open problem to find a complete representation theoretic classification of $G(A^\pm)$ as an exact, nonperturbative function of $\lambda$.

Here, we content ourselves for a particular ansatz of $G(A^\pm)$ motivated by perturbation theory.
Introduce a notion of `pseudo-momentum' $P$ by the exchange algebra:
\bea
P := P(p) \qquad \mbox{where} \qquad G(A^\pm) \Phi := e^{\mp i P(p)} \Phi G(A^\pm) \ .
\eea
Physically, we expect that $P(p)$ is an odd function of $p$ and periodic with periodicity $2\pi$.
This means that the pseudo-momentum $P$ is expandable in a power series of the lattice momentum $p$ as
\bea
e^{iP}=  G(e^{ip}) =e^{ip} e^{i
\sum^\infty_{n=1} b_{2n}( \lambda) \sin n  p}\,. \label{pseudomomentum}
\eea
From general structure of the perturbation theory that dictates hopping of an excitation spin over lattice sites, we expect that
$b_{2n}(\lambda)$ is further expandable as
\be
b_{2n} =\sum^\infty_{\ell = n }   b_{2\ell, 2n} \,\,\lambda^{2 \ell}\,.
\ee

To obey the on-shell conditions $\mathfrak{K} = \mathfrak{K}^* = 0$, one needs to excite two or more spins.
Consider a spin chain with $n$ multiple excitation spins that are asymptotically separated and carry `pseudo-momentum' $P_1, \cdots, P_n$. The momentum central charges now read
\be
&& \mathfrak{K} \,\,\,|\, \Phi^1 \, \Phi^2 \, \Phi^3 \cdots \Phi^n \, \rangle
\, = \, K \,\,|\, \Phi^1 \, \Phi^2 \, \Phi^3 \cdots \Phi^n \, G(A^+)\, \rangle \nonumber \\
&& \mathfrak{K}^* \,|\, \Phi^1 \, \Phi^2 \, \Phi^3 \cdots \Phi^n \, \rangle
= K^* \,|\, \Phi^1 \, \Phi^2 \, \Phi^3 \cdots \Phi^n \,  G(A^-) \,\rangle
\ee
with
\be
K = \sum^n_{k=1} a_k b_k \prod^n_{m=k+1} e^{-iP_m}\,; \qquad
K^* = \sum^n_{k=1} c_k d_k \prod^n_{m=k+1} e^{+iP_m} \, .
\ee
Both should vanish on any on-shell configuration that satisfies
\be
\prod^n_{k=1}e^{-iP_k} =1 \ .  \label{boundarycondition}
\ee
In terms of the lattice momentum, this in general puts the spin chain to obey a version of twisted
boundary condition if closed \footnote{We thank N. Beisert for useful correspondences on issues related to this point.}. Therefore, one might opt to consider the pseudo-momentum defined only in asymptotic limit. Our point is to stress that the condition at quantum level ought to be the one for the `pseudo-momentum' $P$, not for the lattice momentum $p$. As shown in~\cite{Beisert:2005tm}, a unique local solution is given by
\bea
a_k b_k = \alpha(\lambda) (e^{-iP_k}-1)\, ; \, \qquad \, c_k d_k = \beta(\lambda)(e^{iP_k}-1) \, .
\label{alphabeta}
\eea
Note that $\alpha$ and $\beta$ are independent of the lattice site $k$ and the lattice momenta $p_k$'s -- they are parameters {\sl common} to all excitations.

A convenient parametrization of $a$, $b$, $c$ and $d$ are in terms of elliptic variables $X^\pm$ and $h(\lambda), \gamma, f$:
\bea\frac{}{}
&& a =\sqrt{h(\lambda) }\,\,\, \gamma \,, \qquad \qquad b =
-\sqrt{h(\lambda)} {f \over \gamma X^+}(X^+ - X^-)\nonumber\\
&& c  =  \sqrt{h(\lambda)}{i \,\,\, \gamma \over f X^- }\,, \,\, \qquad \,\,\, d =
- \sqrt{h(\lambda)}{i \over \gamma}  (X^+ - X^-) \, .
\eea
With positive definite $h(\lambda)$, defining relations are
\bea
{X^+ \over X^-} = e^{iP}; \qquad \qquad
(X^+)^*=X^- \, ,
\eea
while the shortening condition (\ref{shortening1}) reads:
\bea
X^+ + {1\over X^+}- X^- -{1\over X^-}
={i\over h(\lambda)}\, .
\eea
Explicitly,
\bea
X^\pm = {e^{\pm i{P\over 2}}
\over 4 h \sin{{P\over 2}}}\left(1+ \sqrt{1+16h^2(\lambda)
\sin^2{P\over 2}}\right)\,.
\eea
Unitarity of the representation demands that, modulo a complex phase, $\gamma =\sqrt{ -i (X^+-X^-)}$
and $f=1$. In terms of these variables, the dilatation energy spectrum reads
\bea
E ={h\over i}(X^+- X^-)- {1\over 2}
={1\over 2}\sqrt{1+ 16 h^2(\lambda) \sin^2{P\over 2}}\,\, . \label{energy}
\eea
Classically, the `pseudo-momentum' $P$ is reduced to the lattice momentum $p$, and the elliptic variables $X^\pm$ are reduced to $x^\pm$ where $(x^+/x^-) = e^{i p}$.

As in \cite{Beisert:2005tm}, S-matrices between a pair of magnon excitations is determinable by the $\mathfrak{psu}(2 \vert 2)$ superalgebra. For completeness, we tabulate them in appendix A. Following \cite{janik}, overall scalar phase-factor is also determinable by imposing crossing relations. These S-matrices satisfy unitarity, Yang-Baxter equations and crossing symmetries. From the S-matrices, one
can also construct full-fledged Bethe ansatz equations (BAE). We stress that
all these conclusions are most transparent when the `pseudo-momentum' $P$, equivalently, the elliptic
variables $X^\pm$ are used instead of the lattice momentum $p$, equivalently, $x^\pm$.

A remark is in order. Deformation of integrable spin chain was considered in \cite{Beisert:2005wv}, in which the coupling parameter $\lambda$ is replaced by a set of four deformation parameters. In particular, one of these parameters acts to deform $(x_\pm + \lambda^2 / x_\pm) \rightarrow x_\pm + \sum_{n=3} \alpha_n / (x_\pm)^n$ and hence the energy spectrum as well. It was shown that these deformations are possible while retaining global symmetries of the spin chain intact. Such a deformation appears closely related to the map from the lattice momentum to the pseudo-momentum introduced above. It would be extremely interesting to understand possible relation better.

\section{Spin Chain Hamiltonian from Integrability}

The magnon spectrum expected from the $\mathfrak{psu}(2 \vert 2)$ excitation symmetry and the integrability
is of the form
\be
E = \sqrt{{1 \over 4} + 4\, h^2(\lambda) \sin^2 {p\over 2} }\, .
\label{conjecture}
\ee
It was shown that the energy spectrum receives perturbative corrections only from even-loop orders.
So, the interpolating function $h^2(\lambda)$ is parametrizable as
\bea
h^2(\lambda) = \lambda^2
\sum_{\ell =0}^\infty h_{2 \ell}
\lambda^{2 \ell} \qquad \mbox{with} \qquad h_0 = 1.
\label{interpolatingftn}
\eea
This is the spectrum for the single magnon for each SU(2) sector. Expanding the energy spectrum at weak coupling as a function of lattice momentum $p$,
\bea
E &=&{1\over 2} +   4 h^2(\lambda) \sin^2 {p\over 2} -16 h^4(\lambda) \sin^4 {p\over 2} +\cdots\nonumber\\
&=& \left({1\over 2}\right) +   \Bigl(4 \sin^2 {p\over 2}\Bigr) \lambda^2   +
\Bigl( -16 \sin^4 {p\over 2} + 4 h_2 \sin^2 {p\over 2} \Bigr) \lambda^4
+\cdots \ .  \label{expected}
\eea
As is well-known from planar perturbation theory, $2\ell$-th order contribution to the energy gives rise to lattice shuffling up to $2 \ell$ consecutive sites. At 2-loop order, from $e^{\pm i p}$, the maximal lattice shuffling is for 2 sites. At 4-loop order, from $e^{\pm 2 i p}$, the maximal lattice shuffling is for 4 sites, etc. We shall refer to these as `maximal shuffling' interactions.

At this stage, as shown in the previous section, one needs to bear in mind of the possibility that dynamic spin chain is diagonalized in the pseudo-momentum $P$, not in the lattice momentum $p$. In this case, we need to replace the lattice momentum $p$ in (\ref{conjecture}) by the pseudo-momentum $P$. As the pseudo-momentum is defined as a function of the lattice momentum in perturbation theory, $P = P(p)$, the energy spectrum is still expandable as double series of the `t Hooft coupling and the lattice hopping. Schematically, it takes the form
\bea
E &=& \sum^\infty_{\ell=0}
\lambda^{2\ell}\sum^{\ell}_{n=0} e_{2\ell,2n} \sin^{2n} {p\over 2}
\nonumber\\
&=& \left({1\over 2}\right) +
\Bigl(e_{2,2} \sin^2 {p\over 2}\Bigr) \lambda^2   +
\Bigl( e_{4,2} \sin^2 {p\over 2} +e_{4,4} \sin^4 {p\over 2}\Bigr)
\lambda^4+\cdots
\label{spectrum}
\eea
where $e_{0,0}={1 \over 2}$ counts the classical scaling dimension
 and $e_{2\ell,0}=0 \,\,\,(\ell\,\, \ge \,\, 1)$ is fixed
by the supersymmetry of the ground-state.
The coefficient $e_{2,2}=4$ was computed previously from explicit
computation of quantum dilatation operator at two loop order \cite{Mina, Bak}.

The coefficients $e_{2 \ell, 2n}$ provide a complete information concerning spectra of the quantum dilatation operator, $E(p, \lambda)$. More precisely, from these coefficients, one can determine {\sl both} the generalized marker function $G(e^{ip})$ and the interpolating coupling function $h^2(\lambda)$ as the two functions are mutually independent. Our assertion is based on the observation that, at each order in perturbation theory, the coefficients $e_{2 \ell, 2n}$ in the spectra are in one-to-one correspondence with the two sets of coefficients $(h_{2 \ell}, b_{2 \ell, 2n})$. 

The corresponding integrable Hamiltonian at each order is
well-known.
Consider the odd-site SU(2) chain.
The tree-level part of the Hamiltonian counting the classical scaling dimension
is
\be
H_0 = {1\over 2} \sum^{L-1}_{n=0}
\,\, \mathbb{I} \,.
\ee
The 2-loop part of the Hamiltonian is simply given by~\cite{Mina,Bak}
\be
H_2= \sum^{L-1}_{n=0}(\mathbb{I}- \mathbb{P}_{a_{2n+1} a_{2n+3}})\,,
\ee
where $2L$ is the number of the total sites and $\mathbb{P}$ denotes exchange operator. We consider the asymptotic spin chain where $L$ goes to infinity.
The 4-loop part of the Hamiltonian can be identified as
\be
H_4= {e_{4,2}\over 4} \sum^{L-1}_{n=0}
\, (\mathbb{I}- \mathbb{P}_{a_{2n+1} a_{2n+3}})
+ {e_{4,4}\over 16} \sum^{L-1}_{n=0}[\,(\mathbb{P}_{a_{2n+1} a_{2n+5}}-1) + 4(\mathbb{I}-\mathbb{P}_{a_{2n+1} a_{2n+3}})
]\,,
\label{hamiltonian4}
\ee
where we have used the result of the ${\cal N} = 4$ super Yang-Mills theory~\cite{
Beisert:2003tq,Beisert:2003ys,Eden:2004ua}.
In order to demonstrate that the energy spectrum (\ref{spectrum}) follows from these Hamiltonian, we use the momentum eigenstate of a single $A$-magnon excited on the odd-site SU(2) chain:
\be
| p\rangle_A = \sum_{n=0}^{L-1} e^{i \, n p} |\dots \overbrace{A_2}^{2n+1}\dots\rangle \
\ee
in the background of $A_1$s on the ellipses.
Likewise, for the even-site SU(2) chain, the momentum eigenstate of
a single $B$-magnon is
\be
| p'\rangle_B = \sum_{m=1}^{L} e^{i \, m p'}
|\dots \overbrace{B_2}^{2m} \dots\rangle
\ee
in the background of $B_1$s on the ellipses. Therefore, there are four kinds of excitation states:
\be
| 0\rangle_A \otimes | 0\rangle_B\,, \ \
| p\rangle_A \otimes | 0\rangle_B\,, \ \
| 0\rangle_A \otimes | p'\rangle_B\,,\ \
| p\rangle_A \otimes | p'\rangle_B \,.
\ee
The second and the fourth states are  the momentum $p$ states
of the odd-site SU(2) chain. Below, we shall focus on the odd-site chain
because the odd-site SU(2) chain and even-site SU(2) chain behave
independently for the above quartet of  states.

One easily checks that
\bea
&& H_2\, | p\rangle_A= 4 \sin^2 {p\over 2} \,\,\, | p\rangle_A\nonumber\\
&& H_4 \,| p\rangle_A=
\Bigl( e_{4,2} \sin^2 {p\over 2} + e_{4,4}\sin^4 {p\over 2}\Bigr)\,
| p\rangle_A \, ,
\eea
etc.
In the Hamiltonian (\ref{hamiltonian4}), the maximal shuffling terms $e_{2 \ell, 2 \ell}$ are of particular interest. For example, at 4-loop order, (\ref{expected}) indicates that $e_{4,4}$ equals to $-16$.
In perturbation theory, 
$e_{4,4}$ 
in $H_4$ can be extracted by evaluating anomalous
dimensions of a single $A$-magnon up to 4-loop orders. Since the value
$e_{4,4} = -16$ 
is a consequence of $\mathfrak{psu}(2 \vert 2)$ excitation
symmetry and the integrability, explicit computation would amount to a test of quantum integrability of the ABJM theory.

\begin{figure}[ht!]
\centering \epsfysize=9cm
\includegraphics[scale=0.8]{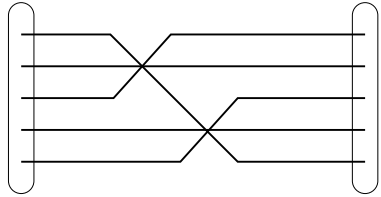}
\caption{\small \sl Maximal shuffling: 4-loop interactions involving five sites. }
\label{fig1}
\end{figure}

\begin{figure}[ht!]
\centering \epsfysize=9cm
\includegraphics[scale=1.0]{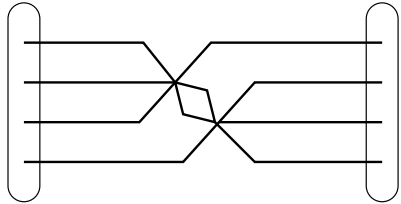}
\caption{\small \sl Next-to-maximal shuffling: 4-loop interaction diagram involving four sites.}
\label{fig2}
\end{figure}

\begin{figure}[ht!]
\centering \epsfysize=9cm
\includegraphics[scale=0.9]{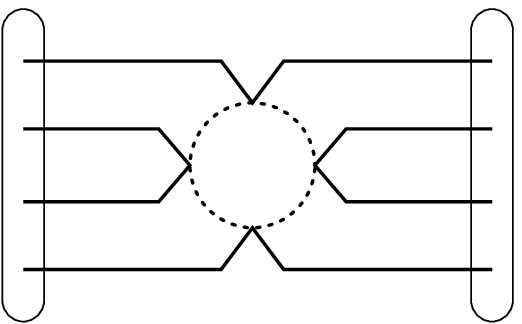}
\caption{\small \sl
Next-to-maximal shuffling: 4-loop interaction diagram involving four sites and fermion
loop.}
\label{figa}
\end{figure}

In the following sections, we shall compute the maximal shuffling term $e_{4,4}$ and the next-to-maximal shuffling term $e_{4,3}$ explicitly and compare with prediction of the $\mathfrak{psu}(2 \vert 2)$ excitation symmetry and the integrability. For the coefficient $e_{4,4}$ of the maximal shuffling term, one needs to compute the diagram in Fig.~\ref{fig1} and its conjugate.
Diagrammatically, there can also arise next-to-maximal shuffling terms. To compute their coefficient $e_{4,3}$, one needs to compute the diagrams
in Fig.~\ref{fig2} and Fig.~\ref{figa}. They give rise to dilatation operators of the form $\sum_n (\mathbb{I} - \mathbb{P}_{n+1,n+3})
(\mathbb{I} - \mathbb{P}_{n+2, n+4})$. In case one type-A and one type-B magnons are excited simultaneously, this operators represent nontrivial interactions between them. This operator, however, is not diagonalized by the state $|\,p\rangle_A \otimes |\,p'\rangle_B$. Absence of such interactions amounts to the statement that $e_{4,3} = 0$.

\section{Quantum Dilatation Operator}
\subsection{Formalism}
We use the anomalous dimension matrix that is defined by
the two-point correlation function:
\be
\langle :\!O(x)\!: \,\,\, :\!O(0)\!:\rangle_{\epsilon}
=  {C^{2L}_\epsilon \over
(x^2)^{{1-\epsilon\over 2} \cdot 2L}
(x^2)^{\gamma(\epsilon)}
}
\ee
where we use the dimensional regularization to control the ultraviolet divergences.
This is related to the collections $A_n$ of  Feynman diagrams at each order defined by
\bea
 \langle :\!O(x)\!: \,\,\, :\!O(0)\!:\rangle_{\epsilon}
&&= (I_\epsilon)^{2L}
e^{-\gamma_\epsilon \ln (x^2\, \Lambda^2 (\epsilon))
}\nonumber\\
&& :=
(I_\epsilon)^{2L} {\rm exp}\,[\,\, \ln(1+ A_2
\lambda^2 + A_4 \lambda^4 +\cdots )\,\,] \ .
\eea
Here, $I_\epsilon$ denotes the dimensionally regularized Euclidean scalar
propagator in the position space:
\be
I_\epsilon = \int {\rmd^{2\omega} p\over (2\pi)^{2\omega}}
{1\over p^2} \, \, e^{i p \cdot x} =
{\Gamma(\omega-1)\over 4\pi^\omega}\,\, {1\over (x^2)^{\omega-1}}
\,,
\ee
with $2\omega= 3-\epsilon$.

From the above, we extract lower loop contributions to the dilatation operator in the ABJM theory
\footnote{For perturbative computations of the dilation operator in ${\cal N}=4$ SYM theory in close parallel to the Feynman diagrammatics we take, see \cite{Gross:2002su}.} by
\bea
&& H_2 = -{\rm lim}_{\epsilon\rightarrow 0}\,\,
\epsilon \,\, A_2\nonumber\\
&& H_4 =
-{\rm lim}_{\epsilon\rightarrow 0}\,\,
2 \epsilon \,\, \left(A_4- {1\over 2} A_2^2\right)\, \nonumber \\
&& H_6 = - {\rm lim}_{\epsilon\rightarrow 0}\,\, 3\epsilon \,\,
\left( A_6 - {1 \over 2} (A_2 A_4 + A_4 A_2) + {1 \over 3} A_2^3 \right).
\label{h6}
\eea
Here, the overall factor $\ell$ in
$H_{2 \ell}$ arises from extracting the dilatation operator as the
charge generating variation of $\ln x^2$
in the ABJM theory.
The leading  singularity in $A_2$ starts from the order $O(\epsilon^{-1})$,  so the leading term in the Laurent expansion of $A_2$ contributes to $H_2$. For the case of $A_4$, the leading singularity in general starts at the $O(\epsilon^{-2})$ order. The coefficient of this leading singularity in $A_4- {1\over 2} A_2^2$ ought to vanish for renormalizability of the theory under consideration. Therefore, the leading singularity starts again from $O(\epsilon^{-1})$ terms. For the case of $A_6$, the leading singularity starts at the $O(\epsilon^{-3})$ order. Again, the coefficient of this leading singularity of $O(\epsilon^{-3})$ and of the next-to-leading singularity $O(\epsilon^{-2})$ vanish for renormalizability. This pattern continues to all higher order contributions to the dilatation operator $H$.

\subsection{Maximal Shuffling and Next-To-Maximal Shuffling Interactions}
We now focus on the ABJM theory. The operator structure involved with Fig.~\ref{fig1} can be constructed
from the operator structure of the two-loop scalar sextet contribution:
\be
O^{\,\, n}_{123}= 2\mathbb{I} - \mathbb{K}_{n+1, n+2}
- \mathbb{K}_{n+2,n+3} +2 \mathbb{P}_{n+1, n+3} \mathbb{K}_{n+1, n+2}
+ 2 \mathbb{P}_{n+1, n+3} \mathbb{K}_{n+2, n+3}- 4 \mathbb{P}_{n+1, n+3}\,,
\ee
where $(\mathbb{K})_{IJ}^{KL}= \delta_{IJ}\delta^{KL}$ refers to contraction operator.
The matrix structure of Fig.~\ref{fig1} is then
\be
O^{\,\,n}_{345} O^{\,\,n}_{123} =
4 \mathbb{I} -8 \mathbb{P}_{n+1, n+3} - 8 \mathbb{P}_{n+3, n+5} + 16
\mathbb{P}_{n+3, n+5} \mathbb{P}_{n+1, n+3} + \cdots
\ee
where the ellipses denotes the omission of the terms involving any types of contraction between the odd and the even sites. Similarly, there is also a diagram corresponding to the conjugate of Fig.~\ref{fig1}. For this, the operator structure takes the form,
\be
 O^{\,\,n}_{123} O^{\,\,n}_{345} =
4 \mathbb{I} -8 \mathbb{P}_{n+1, n+3} - 8 \mathbb{P}_{n+3, n+5} + 16
\mathbb{P}_{n+1, n+3} \mathbb{P}_{n+3, n+5}  + \cdots
\ee
Therefore the full operator structure
becomes
\bea
&& O^{\,\,n}_{12345}= O^{\,\,n}_{123} O^{\,\,n}_{345}+
O^{\,\,n}_{345} O^{\,\,n}_{123}
\nonumber\\
&& =8 \mathbb{I} -16 \mathbb{P}_{n+1, n+3} - 16 \mathbb{P}_{n+3, n+5} + 16
\mathbb{P}_{n+1, n+3} \mathbb{P}_{n+3, n+5}+   16
\mathbb{P}_{n+3, n+5} \mathbb{P}_{n+1, n+3}+ \cdots \ .
\eea
If we put only  $A_a\,\,$/$\,\,B_a$ magnons to
the odd-$\,\,$/$\,\,$ even-sites of the chain,
the above can be rewritten as
\be
O^{\,\, n}_{12345}= 16 \mathbb{P}_{n+1, n+5} - 8 \mathbb{I} + \cdots \ ,
\ee
where we have used the identity
\be
\epsilon_{Ia_1 a_3 a_5}\epsilon_{I b_1b_3 b_5} = \mathbb{I}
 + \delta^{b_3}_{a_1} \delta^{b_5}_{a_3}
\delta^{b_1}_{a_5}+ \delta^{b_5}_{a_1} \delta^{b_1}_{a_3}
\delta^{b_3}_{a_5}- \mathbb{P}_{13}-\mathbb{P}_{35} - \mathbb{P}_{15}=0\,.
\ee
The same operator structure arises from $-{1\over 2}\, A^2_2$, the second-order effects of the two-loop
contribution to the dilatation operator. Hence, adding these two
contributions, one obtains the coefficient $e_{4,4}$.

\begin{figure}[ht!]
\centering \epsfysize=9cm
\includegraphics[scale=0.9]{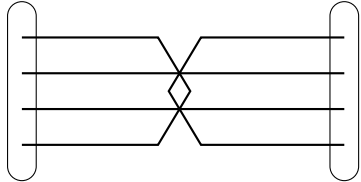}
\caption{\small \sl Remaining
4 loop interaction diagram involving four sites}
\label{fig4}
\end{figure}

The operator structure of the diagrams in Fig.~\ref{fig2}
involves 4-site interactions and can be extracted as
\be
O^{\,\,n}_{1234} &=& O^{\,\,n}_{123} O^{\,\,n}_{234}+ O^{\,\,n}_{234}
O^{\,\,n}_{123}\, \nonumber \\
&=& 32(\mathbb{P}_{n+1,n+3}- \mathbb{I})( \mathbb{P}_{n+2,n+4}- \mathbb{I})+
8(2\mathbb{P}_{n+1,n+3}+2 \mathbb{P}_{n+3,n+5} -3\mathbb{I})+\cdots \,.
\ee
Again, the same operator structure arises from $-{1\over 2}\, A^2_2$
contribution to the dilatation operator. Fermions also contribute to the 4-site operators $\mathbb{P}_{n+1,n+3} \mathbb{P}_{n+2,n+4}$.
This involves fermion loop as depicted in Fig.~\ref{figa}
and the corresponding matrix structure is given by
\be
\bar{O}^{\,\,n}_{1234}= O^{\,\,n}_{1234} + 32 \mathbb{I}
+\cdots \,.
\ee
Operators of the form $(\mathbb{P}_{n+1,n+3}- \mathbb{I})(\mathbb{P}_{n+2,n+4}- \mathbb{I})$
contributes to $e_{4,3}$ in $H_4$. As said, the $\mathfrak{psu}(2 \vert 2)$ excitation symmetry and the integrability predicts that $e_{4,3} = 0$. In the following sections, we shall confirm this explicitly
--- the contributions of Fig.~\ref{fig2}, Fig.~\ref{figa} and second-order effects of 2-loop contributions all cancel one another.

There are many other diagrams that arise at 4-loops and contribute to lower shuffling operators.
Though we shall not evaluate any of these terms in this paper, for completeness of our discussions,
we shall list a class of relevant diagrams. The operator structure of Fig.~\ref{fig4} is proportional to
\be
Q^{\,\,n}_{1234}= 4 \mathbb{P}_{n+1,n+3} + 4 \mathbb{P}_{n+2,n+4}-4 \mathbb{I} + \cdots
\ee
Therefore this 4-loop diagram is relevant to the computation
of the coefficient $e_{4,2}$.
\begin{figure}[hb!]
\centering \epsfysize=9cm
\includegraphics[scale=0.9]{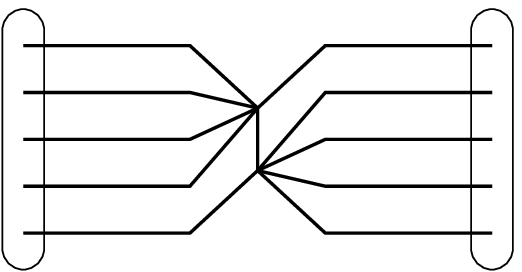}
\caption{\small \sl
4-loop diagram involving 5 sites and
only $\mathbb{K}$ operators
}
\label{figa1}
\end{figure}
\begin{figure}[ht!]
\centering \epsfysize=9cm
\includegraphics[scale=0.9]{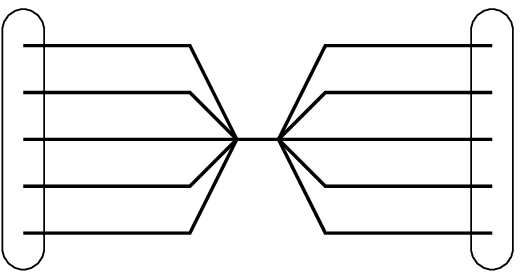}
\caption{\small \sl
4-loop diagram involving 5 sites and
only $\mathbb{K}$ operators
}
\label{figa2}
\end{figure}

\begin{figure}[ht!]
\centering \epsfysize=9cm
\includegraphics[scale=0.9]{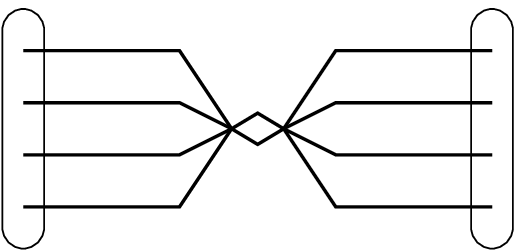}
\caption{\small \sl
4-loop diagram involving 4 sites and
only $\mathbb{K}$ operators
}
\label{figa3}
\end{figure}

\begin{figure}[ht!]
\centering \epsfysize=9cm
\includegraphics[scale=0.9]{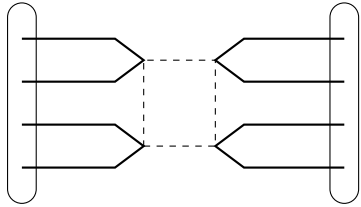}
\caption{\small \sl
4-loop diagram involving 4 sites and fermion loop
}
\label{figc}
\end{figure}

\begin{figure}[ht!]
\centering \epsfysize=9cm
\includegraphics[scale=0.9]{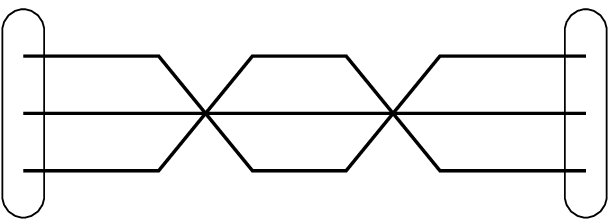}
\caption{\small \sl
4-loop interaction diagram involving 3 sites}
\label{fig10}
\end{figure}

\begin{figure}[ht!]
\centering \epsfysize=9cm
\includegraphics[scale=0.9]{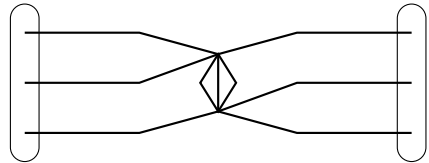}
\caption{\small \sl
Another 4-loop interaction diagram involving 3 sites}
\label{fig9}
\end{figure}
There are many 4-loop diagrams involving only $\mathbb{K}$ operators. Such diagrams involving 5-sites are illustrated in Fig.~\ref{figa1} and Fig.~\ref{figa2} . Further diagrams involving 4-sites are illustrated in
Fig.~\ref{figa3} and Fig.~\ref{figc}.
The 4-loop diagrams contributing operators of 3-sites are proliferated. In Fig.~\ref{fig10} and
Fig.~\ref{fig9}, we illustrate two representative diagrams arising from sextet interaction vertices
only. They contribute to the coefficient $e_{4,2}$.

\section{3-Site and 4-Site Interactions}
In this section, we shall first compute contributions to 3-site operators from 2-loop diagrams and 4-site operators from 4-loop diagrams. As discussed in the last section, the second-order effect of the 2-loop
diagrams will also contribute to $H_4$ along with 4-loop diagrams.

\subsection{2 loops and 3-site interactions}
\begin{figure}[ht!]
\centering \epsfysize=9cm
\includegraphics[scale=0.9]{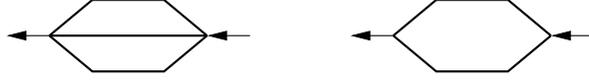}
\caption{\small \sl  Basic
2-loop and 1-loop diagrams with a momentum flow}
\label{fig6}
\end{figure}

For the computation of the 4-loop diagrams,
we begin with the computation of  the 2-loop
and the 1-loop diagrams with a momentum flow $P$.
We first compute the 2-loop contribution of Fig.~\ref{fig6} using the
dimensional regularization. In this diagram, we suppress symmetry factor first and then reinstate it
in the end. We then have
\be
L_3(P)= \int {\rmd^{2\omega} p\over (2\pi)^{2\omega}}
{\rmd^{2\omega} q\over (2\pi)^{2\omega}}
{1\over p^2}  {1\over q^2} {1\over (p+q+P)^2} =
{(\Gamma(\omega -1))^3 \Gamma(3-2\omega) \over
(4\pi)^{2\omega} \Gamma(3\omega-3) (P^2)^{3-2\omega}}\,.
\ee
To get this result, we have used
\be
G(a,b)=
 \!\!\int \!\!{\rmd^{2\omega} p\over (2\pi)^{2\omega}}
{(4\pi)^{\omega} (P^2)^{a+b-\omega} \over (p^2)^a ((p+P)^2)^b} =
{\Gamma(a+b-\omega) \Gamma(\omega-a) \Gamma(\omega-b) \over
 \Gamma(a) \Gamma(b) \Gamma(2\omega-a-b)
}\,.
\ee
Similarly, the 1-loop contribution in Fig.~\ref{fig6} is evaluated as
\be
L_2(P)= \int {\rmd^{2\omega} p\over (2\pi)^{2\omega}}
{1\over p^2}  {1\over (p+P)^2} =
{(\Gamma(\omega -1))^2 \Gamma(2-\omega) \over
(4\pi)^{\omega} \Gamma(2\omega-2) (P^2)^{2-\omega}}\,.
\ee

\begin{figure}[ht!]
\centering \epsfysize=9cm
\includegraphics[scale=0.9]{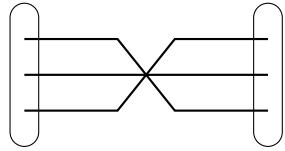}
\caption{\small \sl
3-loop sextet interaction diagram involving 3 sites}
\label{fig7}
\end{figure}
Based on these building blocks, we next compute the diagram of Fig.~\ref{fig7}. It is
proportional to the operator $\sum {O^{\,\,n}_{123}}$:
\be
K_2 =(I_\epsilon)^3 A_2 = (I_\epsilon)^3 (x^2 \pi)^{3-2\omega}\,\,
a_2\,\,
\sum {O^{\,\,n}_{123}} \ ,
\ee
where $a_2$ is given by
\bea
&& a_2 = (-(2\pi)^2)\cdot (I_\epsilon)^{-3} (x^2 \pi)^{2\omega-3}
  \int {\rmd^{2\omega} P\over (2\pi)^{2\omega}}
(L_3(P))^2 e^{iP\cdot x}\nonumber\\
&& \ \ \ \ \, =
-{(2\pi)^2(\Gamma(\omega -1))^3 (\Gamma(3-2\omega))^2 \Gamma(5\omega -6)
\over (4\pi)^3 (\Gamma(3\omega-3))^2 \Gamma(6-4\omega)
} \nonumber\\
&& \ \ \ \ \,
 = -{1\over 4\epsilon}[1+\epsilon (1-\psi(1/2))+O(\epsilon^2)]
\,.
\eea
Here, the polygamma function $\psi(1/2)$ takes the value:
\be
\psi(1/2)= -{\bf C}- 2\ln 2 = -1.963510026\dots
\ee
where ${\bf C}$ denotes the Euler's constant.
For the Fourier transform, we used the formula,
\be
 \int {\rmd^{2\omega} p\over (2\pi)^{2\omega}}
{1\over (p^2)^\alpha} \, \, e^{i p \cdot x} =
{\Gamma(\omega-\alpha)\over 4^\alpha\pi^\omega \Gamma(\alpha)
}\,\, {1\over (x^2)^{\omega-\alpha}}
\,.
\ee
We see that the part of the 2-loop Hamiltonian $H_2$ proportional to $\mathbb{P}_{n+1,n+3}$ is solely
coming from this sextet interaction contribution $A_2$. Therefore, this part of $H_2$ becomes
\be
H^{3s}_2= -{\rm lim}_{\epsilon\rightarrow 0} \,\, \epsilon
\,\, a_2\,\, \sum O^{\,\, n}_{123}
= {1\over 4}\sum O^{\,\, n}_{123} = \sum \Big[
{1\over 2}\,\, \mathbb{I}- \mathbb{P}_{n+1, n+3}+ \cdots \Big]\,,
\ee
This reproduces the previous result obtained in Refs.~\cite{Mina,Bak} .

\subsection{4-loops and absence of 4-site interactions}
We now turn to the coefficient $e_{4,3}$ of the operator $O^{\,\, n}_{1234}$. The contributions
come from the 4-loop diagram of Fig.~\ref{fig2} and also from $-{1\over 2} A_2^2$, the second-order effect of the 2-loop contribution. If the integrability were to hold, contributions to $e_{4,3}$ ought to be zero.

We first compute the 4-loop contribution of Fig.~\ref{fig2}. The full contribution including
the $(I_\epsilon)^4$ part takes the form
\be
K^{4s}_4 = (I_\epsilon)^4 
A^{4s}_4=  (I_\epsilon)^4 (x^2 \pi)^{6-4\epsilon}
\,\, b_4\,\, \sum O^{\,\, n}_{1234} \,.
\ee
To proceed further, we also need the following
double integral:
\be
J_2 \equiv I_5(\,\,1\,\,, 3-2\omega,\,\,1\,\, ,
3-2\omega,2-\omega)\, ,
\ee
where
\bea
I_5(w_1,w_2, w_3,w_4,w_5) =
 \!\!\int\!\! {\rmd^{2\omega} k\over (2\pi)^{2\omega}} {\rmd^{2\omega}l \over (2\pi)^{2\omega}}
{ (4\pi)^{2\omega}\,\,
(p^2)^{-2\omega +
\sum^5_{k=1}w_k}
\over (k^2)^{w_1} (l^2)^{w_2}((l\!-\!p)^2)^{w_3}
((k\!-\!p)^2)^{w_4} ((k\!-\!l)^2)^{w_5}} \ .
\label{twoloop}
\eea
The coefficient
$b_4$ can then  be computed as
\be
b_4= J_2\cdot {(\Gamma(\omega-1))^4\over 4^4 \pi^2} \left[
 \Gamma(3-2\omega)\over \Gamma(3\omega-3)
\right]^2 \cdot {\Gamma(2-\omega)\Gamma(8\omega -10)\over
\Gamma(2\omega-2)\Gamma(10-7\omega)}\,.
\ee
In the Laurent expansion in powers of $\epsilon$, the integral
$J_2$ starts with the $O(\epsilon^0)$ order. Therefore,
it can be expanded as
\be
J_2 = {\beta_1}\Big[ 1+ \beta_2 \epsilon + O(\epsilon^2)
\Big]\,.
\ee
Then $b_4$ has the expansion
\be
b_4= -{ \beta_1 \over 128\pi \epsilon^2}
\Big[ 1+ \big(\beta_2-5 -\psi(1)-2\psi(1/2)\big)
\epsilon + O(\epsilon^2)\Big]\,,
\ee
where the polygamma function $\psi(1)$ is given by
\be
\psi(1)= -{\bf C}= -0.577215\dots
\ee
 On the other hand, the corresponding coefficient
of $O^{\,\,n}_{1234}$
from  $-{1\over 2} A_2^2$  is
 again given by $-{1\over 2} a_2^2$. The leading singularity of order $O(\epsilon^{-2})$
contribution from these two ought to cancel each other since the ABJM theory is renormalizable.
We shall shortly show that the 4-site contribution of Fig.~\ref{figa} starts with the order $O(\epsilon^{-1})$. This then the coefficient $\beta_1$ to
\be
\beta_1= - 4\pi\,.
\ee
We checked this numerically with high precision in the Appendix B.
The corresponding dilatation operator is then given by
\be
H^{b}_4= -{\rm lim}_{\epsilon\rightarrow 0} \,\, 2 \epsilon
\Big[
b_4 -{1\over 2}a_2^2
\Big]\sum O^{\,\,n}_{1234} = -{1\over 16}
\Big[
\beta_2 -7 -\psi(1)
\Big]\sum O^{\,\,n}_{1234}\,.
\ee
We found that
\be
\beta_2= 8 +\psi(1)
\ee
by carrying out the integral numerically. Again, see Appendix B. Therefore, we find scalar loop contributions sum up to yield
\be
H^{b}_4  = -{1\over 16}
\,\sum O^{\,\,n}_{1234}\,.
\ee

\begin{figure}[ht!]
\centering \epsfysize=9cm
\includegraphics[scale=0.9]{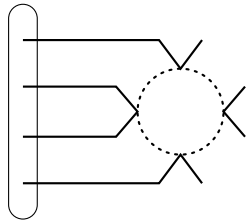}
\caption{\small \sl
4-loop diagram with amputated external lines }
\label{figb}
\end{figure}

Let us now turn to the contribution from the fermion loop contribution, Fig.~\ref{figa}.
There are no more terms that are proportional to $ O^{\,\,n}_{1234}$.
Since this diagram is of $O(\epsilon^{-1})$, we may extract its contribution to the dilatation operator
directly by computing the diagram in Fig.~\ref{figb} with zero momentum flow to the amputated external lines.
The corresponding Feynman integral is given by
\bea
F_4 &=&\int \rmd p \rmd q \rmd k \rmd l {1\over (k-p)^2}  {1\over (k-q)^2}  {1\over (l-p)^2}
 {1\over (l-q)^2}{{\rm tr}\, p \!\!\!\!\slash \, q \!\!\!\!\slash \over p^2 q^2 k^2}
\nonumber\\
&=& \int \rmd p \rmd q \rmd k \rmd l {1\over (k-p)^2}  {1\over (k-q)^2}  {1\over (l-p)^2}
 {1\over (l-q)^2}\left({2\over p^2 k^2}-
{(p-q)^2)\over p^2q^2 k^2}
\right)\,,
\eea
where $\rmd p$ abbreviates for $\rmd^{2\omega} p/(2\pi)^{2\omega}$, etc.
Then, reinstating the symmetry factor, the contribution yields
\be
A_F = F_4\cdot {4!\over 4!}\cdot (2\pi)^4 \cdot {1\over 2}
\,\sum O^{\,\,n}_{1234}
\ee
To compute $F_4$, we perform $l$-integral first, $p$ and $q$
integral next, and finally $k$-integral using the definition $I_5$ and $G(a,b)$.
One finds
\be
F_4 = {1\over (4\pi)^{4\omega}} G(1,1) G(5-3\omega,1)\left[
2 G(2-\omega,1) G(4-2\omega,1)- I_5 (1,1,1,1, 1-\omega)
\right]\,.
\ee
One can evaluate
\bea
&&
G(1,1) G(5-3\omega,1) ={\pi\over \epsilon}(1+ O(\epsilon))
 \nonumber\\
&&
2 G(2-\omega,1) G(4-2\omega,1)= {4\pi\over \epsilon}\left[
1+ \epsilon\,(3+ \psi(1))+ O(\epsilon^2)
\right]\,.
\eea
In the Appendix B, we further find numerically that
\be
 J_3= I_5 (1,1,1,1, 1-\omega)=
{4\pi\over \epsilon}\left[
1+ \epsilon\,(1+ \psi(1))+ O(\epsilon^2)
\right]
\ee
Hence,
\be
A_F = {1\over 2\epsilon}\cdot {1\over 32}
\,\sum \bar{O}^{\,\,n}_{1234}
\ee
and the fermion contribution the Hamiltonian $H_4$ reads
\be
H^f_4 = 2 {d\, A_f\,\over d\ln \Lambda}=  {\rm lim}_{\epsilon\rightarrow 0} 2 \epsilon A_F = {1\over 16}
\,\sum \bar{O}^{\,\,n}_{1234} \,.
\ee
Adding the bosonic and the fermionic contributions, the total 4-site contribution becomes
\be
H^{4s}_4=H^b_4+H^f_4=\sum \left(\,2\,\, \mathbb{I} +\cdots \right)\, ,
\ee
and $e_{4,3} = 0$ identically. We see that this conclusion fits exactly with the $\mathfrak{psu}(2 \vert 2)$ excitation symmetry and the integrability.

\section{5-Site Interactions}
We now turn to the 5-site interaction generated by the diagram
in Fig.~\ref{fig1}. $\!\!\!$\footnote{We are grateful to Joe Minahan pointing out a potential error to the final conclusion in this section concerning (in)compatibility with lattice mommentum.} This computes the coefficient $e_{4,4}$.
Below, the double integral (\ref{twoloop}) for some particular choice of
$(w_1, \cdots, w_5)$ will play a central role. We need to evaluate the integral
\be
J_1 \equiv I_5(2-\omega,3-2\omega, 2-\omega,
3-2\omega,\,\,1\,\,)\,.
\ee
The whole contribution of Fig.~\ref{fig1} then takes the form
\be
K^{5s}_4 = (I_\epsilon)^5 (x^2\, \pi )^{6-4\epsilon}
A^{\rm 5s}_4=  (I_\epsilon)^5 (x^2\, \pi )^{6-4\epsilon}
\,\, a_4\,\, \sum O^{\,\, l}_{12345} \,,
\ee
where
\be
a_4= J_1\cdot {(\Gamma(\omega-1))^5\over 4^4 \pi^2} \left[
{\Gamma(2-\omega) \Gamma(3-2\omega)\over \Gamma(2\omega-2)
\Gamma(3\omega-3)}
\right]^2 \cdot {\Gamma(9\omega -11)\over
\Gamma(11-8\omega)}\,.
\ee

In the Laurent expansion in $\epsilon$, $J_1$ starts with $O(\epsilon^{-1})$. Therefore, the integral
$J_1$ is expandable as
\be
J_1 = {\alpha_1\over \epsilon}\Big[ 1+ \alpha_2 \epsilon + O(\epsilon^2)
\Big]\,.
\ee
Then $a_4$ is expanded as
\be
a_4= -{3 \alpha_1 \pi\over 64\epsilon^2}
\Big[ 1+ \big(\alpha_2-10 -3\psi(1/2)\big) \epsilon + O(\epsilon^2)\Big]\,.
\ee
On the other hand, ${1\over 2} a_2^2$ has the expansion:
\be
{1\over 2} a_2^2= {1 \over 32\epsilon^2}
\Big[ 1+ \big(2 -2\psi(1/2)\big) \epsilon + O(\epsilon^2)\Big]\,.
\label{a22}
\ee
Again, the coefficients of $O(\epsilon^{-2})$
in $a_4 - {1\over 2}a_2^2$ must vanish by the renormalizability of the ABJM theory.
This determines $\alpha_1$ as
\be
\alpha_1= -{2\over 3\pi}\,.
\ee
By numerical integration with high precision, we found that $\alpha_1$ agrees with $ -2/(3\pi)$ with high accuracy. Again, see Appendix B. The corresponding contribution to $H_4$ is then given by
\be
H^{ 5s}_4= -{\rm lim}_{\epsilon\rightarrow 0}\,\, 2 \epsilon\,
\Big[
a_4 -{1\over 2}a_2^2
\Big]\sum O^{\,\,n}_{12345} = -
\Big[
\alpha_2 -12 -\psi(1/2)
\Big]\Big[\sum \mathbb{P}_{n+1,\, n+5}+\cdots \Big]\,.
\ee
The coefficient $\alpha_2$ is computed in the Appendix B as
\be
\alpha_2 = 13+ \psi({1/ 2})\,.
\ee
The $H^{5s}_4$
is thus given  by
\be
H^{5s}_4= - \sum \mathbb{P}_{n+1,\, n+5}+\cdots \ .
\ee
Comparing this with the 5-site maximal shuffling term in $H_4$, we deduce that
\be
e_{4,4} = -16\,. \label{disagreement}
\ee
This agrees precisely with the prediction, $e_{4,4} = -16$. This prediction was based on $\mathfrak{psu}(2 \vert 2)$ excitation symmetry and the integrability. On the other hand, the prediction was also based on a tacit assumption that the magnon excitation is diagonalized by {\sl lattice} momentum eigenstate. Is it
possible that the magnon excitation is actually diagonalized by pseudo-momentum eigenstate? We answer this question in next section.

We close this section with a remark on internal consistency check of the above computations. It is also possible to extract the 6-loop contribution to the Hamiltonian, $H_6$ in (\ref{h6}). There, we first need to make sure that terms of order
$O(\epsilon^{-3})$ and
$O(\epsilon^{-2})$ in $[A_6 - {1 \over 2}(A_4 A_2 + A_2 A_4) + {1 \over 3} A_2^3]$ vanish identically by the renormalizability of the ABJM theory. Cancelation of these terms provide a set of stringent consistency test for the computation of $a_4$ since combinations of $a_4$ and $a_2$ (as well as cubic of $a_2$) ought to cancel against $a_6$. The check is straightforward. We confirmed that these singular terms in $\epsilon$-expansion indeed cancel, ensuring that we computed $a_4$ correctly. Details of this computation, along with explicit evaluation of $H_6$, will be reported in a separate paper.

\section{Interpretation and Discussions}

\subsection{Interpretation}
In the last section, explicit computations showed that the 4-loop dilatation operator shows the structure of anticipated maximal and next-to-maximal shuffling term. Moreover, the coefficient of them matched with the value (\ref{conjecture}) dictated by the $\mathfrak{psu}(2 \vert 2)$ excitation symmetry and integrability.
In this subsection, we compare the result with consideration of pseudo-momentum discussed in section 2. We explain that, as in the ${\cal N}=4$ super Yang-Mills theory, the dilatation operator of the ABJM theory is diagonalized by the lattice momentum eigenstate, not by the pseudo-momentum eigenstate.

To show this, we revisit the energy spectrum (\ref{conjecture}) of a single SU(2) magnon but now assuming that it is an eigenstate of in terms of the pseudo-momentum. Using the relation
(\ref{pseudomomentum})
\bea
P(p) =  p + \sum^\infty_{n=1} \sum^\infty_{\ell=n}   b_{2\ell, 2n} \,\lambda^{2 \ell} \, \sin n p \
\eea
and the interpolating coupling function (\ref{interpolatingftn})
\bea
h^2(\lambda) = \lambda^2
\sum_{\ell = 0}^\infty h_{2\ell} \lambda^{2 \ell} \qquad \mbox{with} \qquad
h_0 := 1 \ , \label{h-function}
\eea
the energy spectrum
\bea
E = \sqrt{ {1 \over 4} + 4\, h^2(\lambda) \sin^2 {P(p)\over 2} }\,,
\ee
is now expanded in terms of the lattice momentum as
\bea
E = {1\over 2} +   \Big[4 \sin^2 {p\over 2} \Bigr] \lambda^2   +
\Bigl[-(16 + 8 b_{2,2}) \sin^4 {p\over 2} + (4h_2 + 8 b_{2,2}) \sin^2 {p\over 2} \Bigr] \lambda^4 + \cdots \ .
\eea
This falls into the pattern we conjectured for the most general spectrum (\ref{spectrum}):
\bea
E &=& \sum^\infty_{\ell=0}
\lambda^{2\ell}\sum^{\ell}_{n=0} e_{2\ell,2n} \sin^{2n} {p\over 2}
\nonumber\\
&=& \left({1\over 2}\right) +
\Bigl(e_{2,2} \sin^2 {p\over 2}\Bigr) \lambda^2   +
\Bigl( e_{4,2} \sin^2 {p\over 2} +e_{4,4} \sin^4 {p\over 2}\Bigr)
\lambda^4+\cdots
\eea
with
\bea
e_{2,2} = 4 \nonumber, \quad e_{4,2} = 4 h_2 + 8b_{2,2}, \quad e_{4,4} = - 16 - 8 b_{2,2}, \cdots.
\label{eb-relation}
\eea
One notes that shift from lattice momentum to pseudo-momentum affects the coefficient of the maximal shuffling term. From the last relation \footnote{The first relation corresponds to the 2-loop results \cite{Mina} \cite{Bak}. The second relation carries useful information regarding the interpolating function $h(\lambda)$ to the first nontrivial order in the weak coupling perturbation theory.}, one learns that our computation (\ref{disagreement}) amounts $b_{2,2} = 0$,
 viz. lattice momentum eigenstate diagonalizes the dilatation operator.

It is easy to see that the identification works uniquely at each order in perturbation theory. The coefficients $e_{2\ell, 2n}$ of conjectured terms in the energy spectrum (\ref{spectrum}) is in one-to-one correspondence with the two sets of coefficients $(h_{2 \ell}, b_{2 \ell, 2n})$. Since the latter two originate from totally different functions, the interpolating coupling function $h^2(\lambda)$ and the generalized marker function $G(z)$, we see that the spectrum $E(p, \lambda)$ determines these two functions uniquely.

\subsection{Discussions}
In this paper, we studied realization of $\mathfrak{psu}(2 \vert 2)$ exciation symmetry and potential integrability in the ABJM theory. The theory is very different from the ${\cal N}=4$ super Yang-Mills theory, only sharing these two aspects. Given this, we raised the possibility that eigenstates (not just eigenvlaues) diagonalizing the dilatation operator are not the lattice momentum basis but the more general pseudo-momentum basis whose functional form $P(p)$ is subject to perturbative corrections. Our result shows the contrary: the ABJM theory is essentially the same the ${\cal N}=4$ SYM theory, at least in these aspects.

Our result confirms that the ABJM theory is quantum integrable up to 4-loop order in weak coupling perturbation theory. We focused on the SU(2)$\times$SU(2) inside SO(6) scalar sector and on maximal shuffling operators in the Hamiltonian, but we expect the integrability extends to the full OSp($6 \vert 4)$ and to all operators straightforwardly, as in \cite{allloop}. We believe our result adds further evidence for exact quantum integrability of the ABJM theory. Below, we discuss implications of our results and issues that deserve further investigation.

\begin{list}{$\bullet$}{}


\item Our consideration in section 2 indicates that notion of pseudo-momentum basis is ubiquitous for the off-shell $\mathfrak{psu}(2 \vert 2)$ superalgebra \cite{Beisert:2005tm}. This superalgebra features both the ${\cal N}=4$ SYM theory and the ABJM theory, but since contents and details of the two theories are very different, precise form of the pseudo-momentum $P(p)$ would differ for the two theories. Nevertheless, result of this paper indicates that the dilatation operators of both theories are diagonalized by lattice momentum eigenbasis. The fact that change to pseudo-momentum basis did not actually take place in both theories poses a puzzle and additional hidden structure yet to be understood better. A possible checkpoint would be a direct computation of the off-shell central charges $\mathfrak{K}, \mathfrak{K}^*$ in weak coupling perturbation theory. By comparing the result with the local solution (\ref{alphabeta}), one may be able to understand why all $b_{2\ell, 2n}$ vanish in these two theories.

\item Our result also bears an implication to the interpolating function $h^2(\lambda)$. The leading perturbative correction $h_2$ in (\ref{h-function}) is extractable, for instance, from 3-site shuffling term $e_{4,2}$ in the 4-loop spectrum~\footnote{The computation is currently under progress by J. Minahan, O. Ohlson Sax and C. Sieg (private communications). See also \cite{minahan}.}. From (\ref{eb-relation}), one sees that $e_{4,2}$ is directly related to $h_2$ {\sl if} $b_{2,2}$ vanishes and the dilatation operator is diagonalized by the lattice momentum eigenstates.

\item The notion of pseudo-momentum also calls for revisiting the interpretation of the giant magnon at strong `t Hooft coupling limit for both AdS$_5$ \cite{gm} and AdS$_4$ \cite{gm4} cases.. In the light-cone gauge of the {\sl Lorentzian} string worldsheet, it is always possible to reparametrize each chiral worldsheet coordinates separately. This implies that there can in general be an arbitrariness in identifying the worldsheet momentum conjugate to the chiral worldsheet coordinate with the angular separation of the string configuration in the giant magnon.


\item Extension of our result to the parity-violating ${\cal N}=6$ superconformal Chern-Simons theory with gauge group U$(M)\times$U$(N)$ is straightforward. Two-loop Hamiltonian was constructed in \cite{Bak2} and was found integrable. Our result in this paper indicates that the integrability persists up to four loops: one only needs to replace $\lambda^2$ to the geometric mean of $\lambda^2(M)$ and $\lambda^2(N)$.

\item By introducing a variety of D-branes wrapping supersymmetric cycles in $\mathbb{CP}^3$, one can construct open spin chains. For AdS$_5$ / SYM$_4$ correspondence, several open spin chains including those ending on giant gravitons were found quantum integrable \cite{reflectinggm}. The $\mathfrak{psu}(2 \vert 2)$ superalgebra or appropriate subalgebra continued to act as the excitation symmetry. Similarly, we expect that integrable open spin chains exist in ABJ(M) theories by adding D-branes and flavors \cite{flavor}. With concrete realization and identification of excitation symmetries therein, we expect that one can learn more not only about the quantum integrability but also aspects of the pseudo-momentum in the presence of boundaries.

\item Extension to higher loop orders would be highly desirable. It will teach us not only the magnon energy spectrum and quantum integrability but also other pertinent issues including wrapping interactions, structural (dis)similarity of the dilatation operator with the ${\cal N}=4$ SYM theory etc.

\end{list}

We are currently investigating these issues and intend to report further results in separate publications.

\section*{Acknowledgement}
We are grateful to Niklas Beisert, David J. Gross, Joe Minahan, Didina Serban and Matthias Staudacher
for very helpful discussions and correspondences. SJR thanks warm hospitality of the Kavli Institute for Theoretical Physics during this work. This work was supported in part by 
by Grants from the Korea National Science Foundation R01-2008-000-10656-0, SRC-CQUeST-R11-2005-021, 2008-313-C00175, 2005-084-C00003, 2009-008-0372, EU-FP Marie Curie Research \& Training Networks HPRN-CT-2006-035863 (2009-06318) and the U.S. National Science Foundation under Grant No. PHY05-51164 at KITP.

\appendix

\section{$\mathfrak{psu}(2 \vert 2)$ S-matrices with Pseudo-Momentum}

In this section we record the $SU(2|2)$ invariant S matrix.
Only change is to replace $x^\pm$ of S matrix
in \cite{Beisert:2005tm} by
$X^\pm$ which have more complicated momentum
dependence.
The S matrix reads\footnote{More precisely
one needs to include the correction in
Ref.\cite{Arutyunov:2006yd}.}
\bea
&& {\cal S}_{12}|\phi_1^a\phi^b_2\rangle
= A_{12}|\phi_2^{\{a}\phi^{b\}}_1\rangle
+B_{12}|\phi_2^{[a}\phi^{b]}_1\rangle
+{1\over 2} C_{12}
\epsilon^{ab}\epsilon_{\alpha\beta}|\psi_2^{\alpha}\psi^{\beta}_1
\bar{G}(A^-)\rangle
\nonumber\\
&& {\cal S}_{12}|\psi_1^\alpha\psi^\beta_2\rangle
= D_{12}|\psi_2^{\{\alpha}\phi^{\beta\}}_1\rangle
+E_{12}|\psi_2^{[\alpha}\psi^{\beta]}_1\rangle
+{1\over 2} F_{12}
\epsilon^{\alpha\beta}\epsilon_{ab}|\phi_2^{a}\phi^{b}_1
{G}(A^+)\rangle
\nonumber\\
&& {\cal S}_{12}|\phi_1^a\psi^\beta_2\rangle
= G_{12}|\psi_2^{\beta}\phi^{a}_1\rangle
+H_{12}|\phi_2^{a}\psi^{\beta}_1\rangle
\nonumber\\
&& {\cal S}_{12}|\psi_1^\alpha\phi^b_2\rangle
= K_{12}|\psi_2^{\alpha}\phi^{b}_1\rangle
+L_{12}|\phi_2^{b}\psi^{\alpha}_1\rangle
\eea
where
\bea
&& A_{12}= S^0_{12} {X_2^+ - X_1^-\over X_2^- - X_1^+}\nonumber\\
&& B_{12}= S^0_{12} {X_2^+ - X_1^-\over X_2^- - X_1^+}\left(
1- 2{1-1/(X_2^- X_1^+)\over
1-1/(X_2^- X_1^-)
}\,\,{X_2^+ - X_1^+\over X_2^+ - X_1^-}
\right)
\nonumber\\
&& C_{12}= S^0_{12} {2\gamma_1\gamma_2/(X_2^- X_1^-)\over
1-1/(X_2^- X_1^-)
}\,\, {X_2^+ - X_1^+\over X_2^- - X_1^+}
\nonumber\\
&& D_{12}= -S^0_{12}
\nonumber\\
&& E_{12}= -S^0_{12} \left(
1- 2{1-1/(X_2^+ X_1^-)\over
1-1/(X_2^+ X_1^+)
}\,\,{X_2^- - X_1^-\over X_2^- - X_1^+}
\right)
\nonumber\\
&& F_{12}= S^0_{12} {2\gamma_1\gamma_2/(X_2^+ X_1^+)\over
1-1/(X_2^+ X_1^+)
}\,\, {X_2^- - X_1^-\over X_2^- - X_1^+}
\nonumber\\
&& G_{12}= S^0_{12} {X_2^+ - X_1^+\over X_2^- - X_1^+}\nonumber\\
&& H_{12}= S^0_{12}{\gamma_1\over \gamma_2} \,\,
{X_2^+ - X_2^-\over X_2^- - X_1^+}\nonumber\\
&& K_{12}= S^0_{12}{\gamma_2\over \gamma_1} \,\,
{X_1^+ - X_1^-\over X_2^- - X_1^+}\nonumber\\
&& L_{12}= S^0_{12} {X_2^- - X_1^-\over X_2^- - X_1^+}\,.
\eea
The phase $S_{12}^0$ satisfying
$S^0_{12} S^0_{21}=1$ will be specified below.
The above is for the odd chain S matrices. For the even chain one
can have   $2|2$ excitations transforming under the
$su(2|2)$ symmetriy.
The S matrix for the $N=6$ Chern-Simons theory is consisting of
\bea
&& S^{oo} =S^{ee}= S_{12}^{0} \hat{S}\nonumber\\
&& S^{oe} =S^{eo}= \tilde{S}_{12}^{0} \hat{S}\,.
\eea
where $\hat{S}$ denotes the matrix part of the above ${\cal S}$ matrix
without the phase  $S^0_{12}$.
One also finds the crossing symmetric phases by
\bea
&& S^0_{12} ={1-1/(X^+_2 X^-_1)\over 1-1/(X^-_2 X^+_1)}
\sigma(X_1, X_2)\nonumber\\
&& \tilde{S}^0_{12}=
{X^-_2- X^+_1\over X^+_2 - X^-_1} \sigma(X_1, X_2)
\eea
where $\sigma(X_1, X_2)$ is the dressing phase found
in \cite{Beisert:2006ib}.

The full $S$ matrices above satisfy the unitarity condition
\be
{\cal S}^{f}_{12}  {\cal S}^{f}_{21}= I
\ee
and the Yang-Baxter equation
\be
{\cal S}^{f}_{12}  {\cal S}^{f}_{13}{\cal S}^{f}_{23}
={\cal S}^{f}_{23}  {\cal S}^{f}_{13}  {\cal S}^{f}_{12}\,.
\ee
We note that these structures relies only on local exchange algebras, independent global boundary conditions such as (\ref{boundarycondition})

\vskip0.5cm

\section{Numerical Evaluation of the integrals $J_1$, $J_2$ and $J_3$}
In this appendix, we evaluate the two-loop integrals $I_5(w1,w2,w3,w4,w5)$
in (\ref{twoloop}). It is difficult to find an analytically closed form of them.
However there exist several packages with which we can numerically evaluate them.
 It turns out that the Mathematica packages,
MB\cite{Czakon:2005rk,Smirnov:2009up} and
AMBRE\cite{Gluza:2007rt}
are useful.
 Since these packages utilize a method based on the Mellin-Barnes representation of
integrals, we cast the integrals into the form:
\begin{eqnarray}
I_5(w1,w2,w3,w4,w5)=\int\frac{d z_1}{2\pi i}\int
\frac{d z_2}{2\pi i}\frac{\Gamma(-z_1)\Gamma(\omega-w_{25}-z_1)
\Gamma(\omega-w_{1}+z_1)}{\Gamma(w_1-z_1)}
\\ \frac{\frac{}{}}{}\nonumber
\times\frac{\Gamma(-z_2)\Gamma(\omega-w_{35}-z_2)
\Gamma(\omega-w_4+z_2)}{\Gamma(w_4-z_2)}\\  \nonumber
\times \frac{\Gamma(-\omega+w_{14}-z_1-z_2)
\Gamma(-\omega+w_{235}+z_1+z_2)\Gamma(w_{5}+z_1+z_2)}{
\Gamma(w_2)\Gamma(w_3)\Gamma(w_5)\Gamma(w_{235})
\Gamma(2\omega-w_{235})\Gamma(2\omega-w_{14}+z_1+z_2)},
\end{eqnarray}
where $w_{235}$ denotes $w_2+w_3+w_5$.
Note that this corresponds to a three dimensional
version of the expression (25) in \cite{Bierenbaum:2003ud}. After  modifying
the package AMBRE to be applicable to three dimensions, a direct
application of these packages produces the numerical value of the desired function
\begin{eqnarray}
J_1&=&I_5(2-\omega,3-2\omega,2-\omega,3-2\omega,1)=
\frac{\alpha_1}{\epsilon}\left[1+\alpha_2\epsilon+O(\epsilon^2)\right]\\ \nonumber
&=&- \frac{0.212207\ldots}{\epsilon}\left[1+(11.0365\ldots)\epsilon +O(\epsilon^2)\right]
\end{eqnarray}
From this, one may easily read $\alpha_1=-\frac{2}{3\pi }$
and $\alpha_2=13+\psi(\frac{1}{2})$.
For the integral $J_2$, we have found the following numerical value
 \begin{eqnarray}
J_2&=&I_5(1,3-2\omega,1,3-2\omega,2-\omega)=
\beta_1\left[1+\beta_2\epsilon+O(\epsilon^2)\right]\\ \nonumber
&=& -(12.5664\ldots) \left[1+ (7.42278\ldots)\epsilon+O(\epsilon^2)\right] .
\end{eqnarray}
This is consistent with the expressions $\beta_1=-4\pi$
and $\beta_2=8+\psi(1)$. The last integral is evaluated as
 \begin{eqnarray}
J_3&=&I_5(1,1,1,1,1-\omega)= 
\frac{12.5664
\ldots}{\epsilon} \left[1+ (0.422784\ldots)
\epsilon+O(\epsilon^2)\right] \\ \nonumber
&=& {4\pi\over\epsilon} \left[1+ (1+\psi(1))\epsilon+O(\epsilon^2)\right].
\end{eqnarray}


\end{document}